\documentclass[aps,twocolumn,groupedaddress]{revtex4}
\usepackage{epsf} 

\begin{document}

\preprint{ASC SMRG \#690}

\title{Core pinning by intragranular nanoprecipitates in polycrystalline MgCNi$_3$}


\author{L.D. Cooley, X. Song, J. Jiang, and D.C. Larbalestier}
\email[]{ldcooley@facstaff.wisc.edu}
\homepage[]{asc.wisc.edu}
\affiliation{Applied Superconductivity Center, University of Wisconsin, Madison, Wisconsin}

\author{T. He, K.A. Regan, and R.J. Cava}
\affiliation{Department of Chemistry and Princeton Materials Institute, Princeton University, Princeton, New Jersey}

\date{\today}

\begin{abstract}
The nanostructure and magnetic properties of polycrystalline MgCNi$_3$ were studied by x-ray diffraction, electron microscopy, and vibrating sample magnetometry.  While the bulk flux-pinning force curve $F_p(H)$ indicates the expected grain-boundary pinning mechanism just below $T_c \approx 7.2$~K, a systematic change to pinning by a nanometer-scale distribution of core pinning sites is indicated by a shift of $F_p(H)$ with decreasing temperature.  The lack of scaling of $F_p(H)$ suggests the presence of 10 to 20\% of nonsuperconducting regions inside the grains, which are smaller than the diameter of fluxon cores $2\xi$ at high temperature and become effective with decreasing temperature when $\xi(T)$ approaches the nanostructural scale.  Transmission electron microscopy revealed cubic and graphite nanoprecipitates with 2 to 5~nm size, consistent with the above hypothesis since $\xi(0) \approx 6$~nm.  High critical current densities, more than $10^6$~A/cm$^2$ at 1~T and 4.2~K, were obtained for grain colonies separated by carbon.  Dirty-limit behavior seen in previous studies may be tied to electron scattering by the precipitates, indicating the possibility that strong core pinning might be combined with a technologically useful upper critical field if versions of MgCNi$_3$ with higher $T_c$ can be found.
\end{abstract}

\maketitle


The recent discovery \cite{He} of superconductivity above 7~K in MgCNi$_3$ suggests the possibility of a new family of nickel-carbide superconductors.  MgCNi$_3$ is the cubic cousin of layered RNi2B2C compounds (R = rare earth), which exhibit superconductivity up to 23 K for YNi2B2C \cite{Cava94}. It has been proposed that in analogy to copper and bismuth-oxide perovskites, the conduction in MgCNi$_3$ involves holes in the Ni $d$ states \cite{He}, which are usually responsible for magnetism.  The possibility, therefore, of unconventional superconductivity has attracted great interest in the band structure and the physics of the pairing mechanism \cite{Dugale,Shim,Singh,Rosner}.  From an application point of view, it is interesting that MgCNi$_3$ behaves like a dirty-limit intermetallic superconductor, with a high resistivity and a steep slope of the upper critical field $H_{c2}$ at the critical temperature $T_c$ \cite{Li}.  This produces $\mu_0 H_{c2}(0)$ values close to 10~T \cite{Li}, comparable to the reported 8-15 T values for the layered, clean-limit borocarbide compounds \cite{Lan} that have 2 to 3 times higher $T_c$.

In this Letter, we present evidence for nanometer-scale precipitates in MgCNi$_3$, which may explain why dirty-limit behavior is observed.  Bulk pinning-force curves $F_p(H)$ derived from magnetization data show a systematic change with decreasing temperature, from behavior characteristic of grain-boundary pinning near $T_c \approx 7.2$~K to behavior characteristic of core pinning at 1.8~K.  This implies the existence of intragranular pinning sites that are invisible to flux lines at high temperature, when the flux-line diameter $2\xi$ is large, but become effective at low temperature, when $\xi(T)$ is slightly larger than $\xi(0)\approx 6$~nm.  Evidence for both graphite and a cubic phase is provided by high-resolution transmission electron microscopy (TEM), where both phases have 2 to 5~nm size.  These nanoprecipitates probably result from the processing of MgCNi$_3$, since excess carbon is needed to get the highest values of $T_c$ \cite{He}.  MgCNi$_3$ may thus be unique among intermetallic superconductors, since it can be made with intragranular core-pinning sites that are necessary for strong pinning at high fields, a feature not found in e.g. Nb$_3$Sn.  Moreover, if analogs of this compound with higher $T_c$ retain this nanostructure, they should exhibit high upper critical field, due to the electron scattering by the nanostructure, in addition to strong core pinning.  Such a favorable combination is presently found only in Nb-Ti alloy superconductors \cite{CMb}, which are the mainstay of superconducting magnet technology.

\begin{figure}
\epsffile{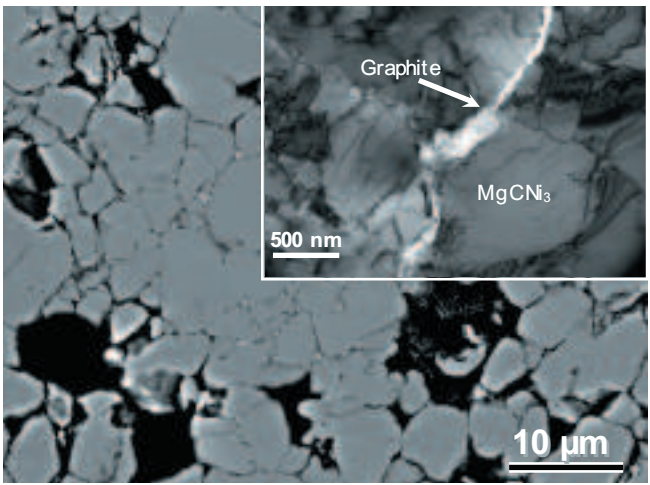}
\caption{Electron microscopy of a polished sample.  MgCNi$_3$ colonies appear as light gray regions in scanning electron microscopy, with excess graphite appearing as the black regions.   Inset: Transmission electron microscopy shows that the MgCNi$_3$ colonies consist of 100 to 300~nm grains, and are separated by a graphite layer.}
\label{fig-semtem}
\end{figure}

A MgCNi$_3$ pellet from a batch reported in \cite{He} was studied, with nominal formula MgC$_{1.5}$Ni$_3$.  Light microscopy of polished pieces cut from the pellet showed a dense, shiny phase with $< 5$\% porosity.  Small regions of pure carbon (as graphite) were visible and indicated by x-ray analyses.  Samples for magnetization and transmission electron microscopy (TEM) characterization were cut from the interior of the pellet.
Scanning and transmission electron microscopy, fig.~\ref{fig-semtem}, show that the samples consist of colonies of fine grains.  SEM analyses show colonies ranging from 1 to 10~$\mu$m in size, with thin regions of carbon in the boundaries between colonies.  TEM analyses show that the grain size is 100-300~nm, typical of a polycrystalline intermetallic compound.  Grain-boundary dislocations were evident along many grain boundaries, suggesting that a substantial number of grains have low angles of misorientation to each other within each colony.  

\begin{figure}
\epsffile{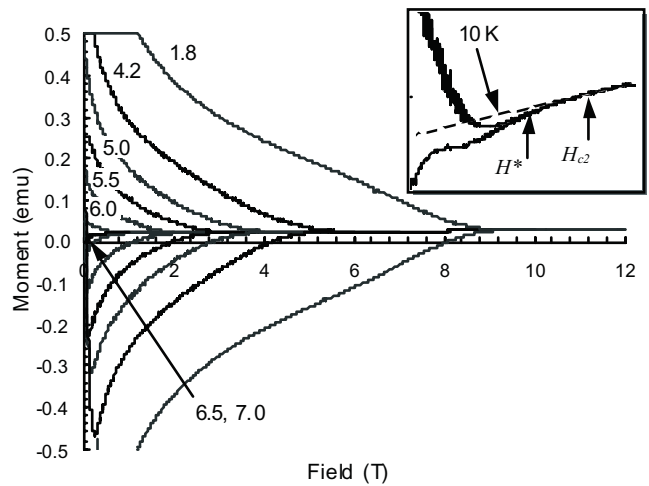}
\caption{Magnetization curves from 1.8 to 10~K.  Inset: 4.2~K data, showing how values for $H^\ast$ and $H_{c2}$ were determined.}
\label{fig-magn}
\end{figure}

Electromagnetic characterization was performed on a 3~mm $\times$ 3~mm $\times$ 0.5~mm prism.  A vibrating sample magnetometer (VSM) was used to measure the sample moment $m(H,T)$ from 1.8 to 330~K and in a field $H$ from 0 to 14~T.  In the superconducting state, raw hysteresis loops for 1.8 to 7.0~K generally were symmetric around a weakly ferromagnetic background measured at 10~K, shown in fig.~\ref{fig-magn}.  A slight bulge is evident at high field for the 1.8~K curve, in comparison to curves at higher temperatures.  The hysteresis loops actually close slightly below the background magnetization, at the irreversibility field $H^\ast(T)$, as shown in fig.~\ref{fig-magn} inset.  A small, reversible diamagnetic magnetic moment (relative to the 10~K background) was visible above $H^\ast$, which changed slope to overlap the 10~K curve at the upper critical field $H_{c2}(T)$.  Some inhomogeneity of the samples was present, which gives an uncertainty of 0.1~T to the $\mu_0 H^\ast(T)$ and $\mu_0 H_{c2}(T)$ values.  The critical field results are similar to the recent results of Li et al. \cite{Li}, namely a slope $\mu_0 dH_{c2}/dT$ of $-1.86$~T/K at $T_c$ and an extrapolated $\mu_0 H_{c2}(0) = 9.2$~T.  Based on the $H_{c2}(0)$ value, the coherence length at zero kelvin $\xi(0) = (2\pi\mu_0 H_{c2})^{1/2}$ is 5.9~nm.

The critical current density $J_c(H,T)$ was determined by applying the standard critical state expression for a thin square prism to the magnetization data \cite{Evetts}, $J_c = 3\Delta M/a$, where $\Delta M$ is the full width of the magnetization hysteresis and $a$ is the half-width of the sample.  From this evaluation, $J_c$ is about $10^3$ to $10^4$~A/cm$^2$ at 4.2~K, assuming the current flows around the entire sample.  However, since the microstructural analyses (fig.~\ref{fig-semtem}) show carbon between grain clusters, the cluster size $\sim\!\!10$~$\mu$m is a better estimate of the current-loop diameter.  Using this length scale, $J_c$ is $1.6 \times 10^6$~A/cm$^2$ at 1~T, 4.2~K, where the hysteresis in magnetic moment is approximately $5 \times 10^{-4}$~A$\cdot$m$^2$ (0.5~emu).

\begin{figure}
\epsffile{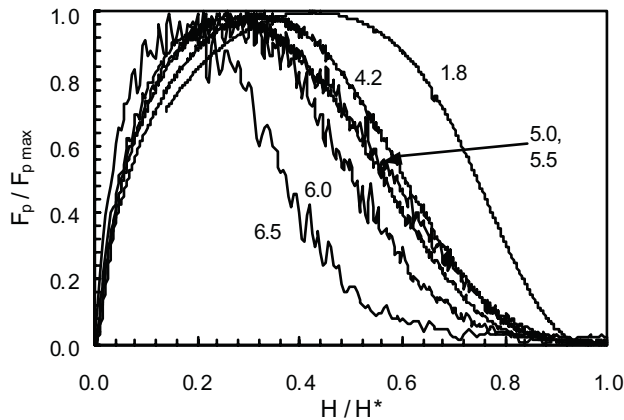}
\caption{Reduced bulk pinning-force curves at 1.8 to 6.5~K.  Note that curves for 5.0 and 5.5~K nearly overlap.}
\label{fig-Fp}
\end{figure}

In conventional superconductors, the flux pinning mechanism associated with microstructural defects is often assessed by analyzing the shape of the bulk pinning-force curve $F_p(H) = \mu_0 H \cdot J_c(H,T)$ as a function of temperature.  Reduced curves $F_p(h,T) / F_{p max}(h,T)$ plotted against a normalized field $h = H/H^\ast(T)$ typically overlap when a single pinning mechanism and pinning center is dominant \cite{FietzWebb}, and this behavior would be expected if e.g.\ grain boundaries alone were the pinning centers.  Such scaling behavior is commonly observed in intermetallic low-temperature superconductors such as Nb$_3$Sn \cite{Hampshire,Kahan,DewHughes}.  By contrast, a constant shape of the bulk pinning-force curve is not obeyed as a function of temperature in the present experiment, as shown in fig.~\ref{fig-Fp}.  Instead, a systematic shift of the reduced pinning-force curve peak toward higher field occurs as the temperature is reduced.  This behavior strongly suggests that different pinning mechanisms and pinning centers are at work for temperatures near to and far below $T_c$, respectively.  It is very unlikely that pinning by grain boundaries is the dominant pinning mechanism at 1.8~K, because the observed grain size of 100 to 300~nm is comparable to that seen in widely-studied Nb$_3$Sn composites, for which the peak of $F_p(H)$ is rarely higher than $h = 0.25$ \cite{Kahan}.  Instead, the 1.8~K curve has the shape expected for pinning by small voids or precipitates \cite{CMb}.  A grain-boundary pinning mechanism is consistent with the 6.5~K curve in fig.~\ref{fig-Fp}.

\begin{figure}
\epsffile{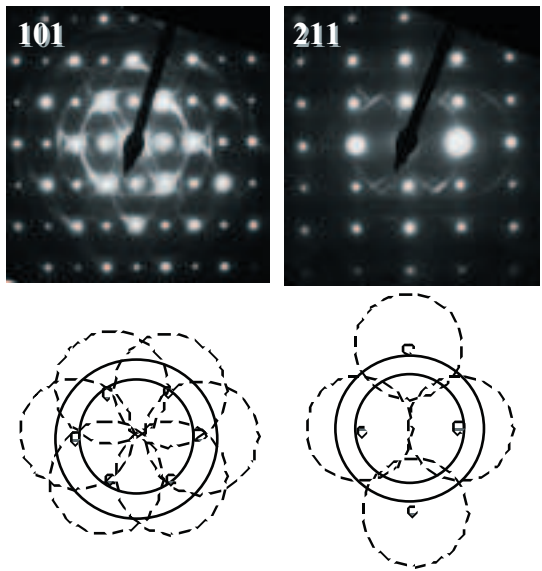}
\caption{Selected area diffraction patterns from the 101 and 211 zone axes, respectively.  Arcs centered around the central electron beam (solid lines in the sketches below) are due to primary diffraction of graphite.  Arcs centered around MgCNi$_3$ diffraction spots (dashed lines in the sketches) are due to double diffraction.}
\label{fig-graphite}
\end{figure}

The presence of nanoprecipitates below the resolution of fig.~\ref{fig-semtem} is revealed by high-resolution transmission electron microscopy.  Fig.~\ref{fig-graphite} shows selected-area diffraction patterns indicating pure graphite nanoprecipitates.  The rectangular array of spots is produced by the MgCNi$_3$ for the indicated zone axes.  Graphite is indicated by the weak arcs centered around the incident electron beam, where the 101 (0.208 nm) and 015 (0.146 nm) arcs are brightest.  Double diffraction produces arcs centered around the parent MgCNi$_3$ diffraction spots. The appearance of arcs rather than rings indicates that the graphite is textured with respect to the MgCNi$_3$ lattice.  The graphite nanoparticles were 2-5 nm in size.

\begin{figure}
\epsffile{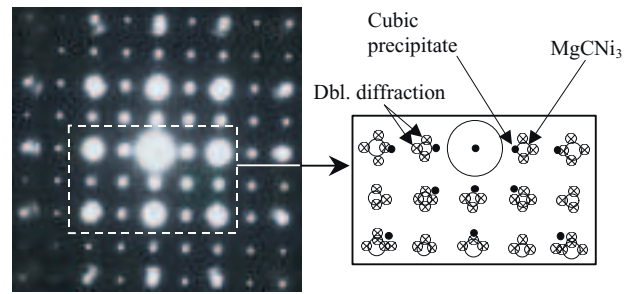}
\caption{Selected area diffraction pattern from the 100 zone axis for a different region than in fig.~\ref{fig-graphite}.  The sketch indicates primary diffraction spots from MgCNi$_3$ ($\bigcirc$) and the cubic precipitate ($\bullet$), such as the 100 spots identified by arrows.  Double diffraction spots are also indicated ($\otimes$), where primary diffraction beams from MgCNi$_3$ serves as incident beams for the precipitate.  These double diffraction spots are blurred with the primary spots to produce the overlap seen.}
\label{fig-cubic}
\end{figure}

\begin{figure}
\epsffile{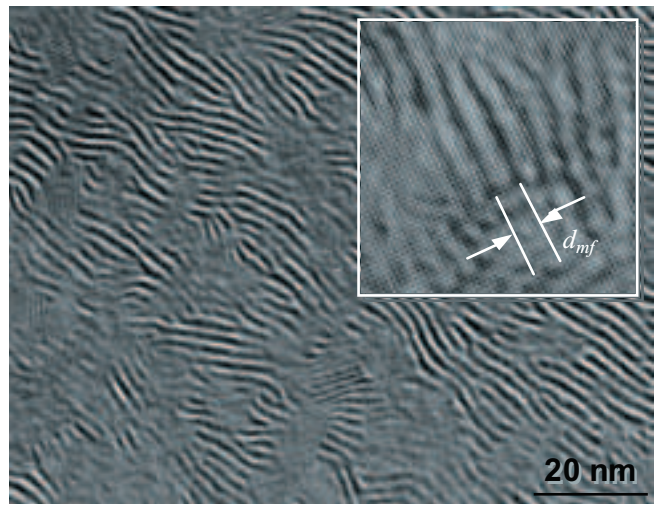}
\caption{High-resolution TEM images from another, thin region of the sample, taken along the 301 zone axis showing the size of nano-particles.  Inset: High resolution image showing moiré fringe spacing relative to atomic rows of the MgCNi$_3$ phase.  The lattice fringe separation spans approximately 5 MgCNi$_3$ unit cells.}
\label{fig-moire}
\end{figure}

Figs.~\ref{fig-cubic} and \ref{fig-moire} show evidence for a cubic precipitate. Diffraction patterns taken from various zone axes, shown in fig.~\ref{fig-cubic}, have secondary spots in addition to the primary spots from MgCNi$_3$.  Tertiary spots caused by double diffraction can also be seen.  All of the secondary spots follow the cubic symmetry of the main MgCNi$_3$ phase. Using the lattice parameter of MgCNi$_3$, $d_0 = 0.381$~nm \cite{He} as an internal reference,  the cubic precipitate has a lattice parameter of about 0.47~nm based on the diffraction pattern.  Fig.~\ref{fig-cubic} also indicates that the cubic precipitates have cube-on-cube texture with the MgCNi$_3$ matrix.  Since the precipitate lattice parameter is 25\% larger than that of the matrix, there is close to a 5-on-4 commensuration between the two phases.  Consistent with this analysis, moiré fringes dominate high-resolution TEM images, such as those shown in fig.~\ref{fig-moire}.  From the fringe spacing $d_{mf}$ of 1.96~nm, a lattice parameter of $d_1 = 0.47$~nm can be obtained from the reciprocal relationship $d_{mf} = d_0 d_1(d_1 –- d_0)^{-1}$.  This agrees with the value obtained from diffraction above.  Mismatch strains are absent in fig.~\ref{fig-moire} probably due to the commensuration.  Domains $\sim\!\!5$~nm in size are also evident, indicating the nanoprecipitate size.

To explain the unusual flux-pinning behavior, we note that the plot in fig.~\ref{fig-Fp} is strikingly similar to the behavior reported in \cite{CMb} for optimized Nb48wt.\%Ti strands.  In that system, strong pinning was produced by a very fine 2-phase nanostructure of 1 to 4~nm thick, nearly pure Ti precipitates separated by 5 to 10~nm in a matrix of superconducting Nb-Ti.  Since the (nonsuperconducting) precipitate thickness was not too small compared with the flux-line core diameter $2\xi \approx 10$~nm at 4.2~K, and since there was a higher number of precipitates than flux lines, the authors reasoned that individual core-pinning interactions could be summed up directly to give a pinning-force curve with a peak at $h \approx 0.5$.  However, since just below $T_c = 9.1$~K the fluxon core was much larger, due to the divergence of $\xi(T) \propto (1 -– T/T_c)^{-1/2}$, the nanometer-scale variations became invisible to the flux lines and larger scale variations then became the pinning sites.  This produced a pinning-force curve with a peak below $h = 0.25$ at 8.5~K.  The lack of temperature scaling of $F_p(H)$ in \cite{CMb} thus resulted from the combination of having nonsuperconducting regions distributed on a scale less than $\xi(0)$ and the diverging temperature dependence of $\xi$ as $T \rightarrow T_c$.

In the present experiment, a similar distribution of pinning sites at length scales comparable to $\xi(0) \approx 6$~nm is present.  We surmise, therefore, that the shift of the pinning force curve in fig.~\ref{fig-Fp} is the result of flux pinning by the nanoprecipitates at low temperature and its absence at $T$ near $T_c$.  Graphite is not a superconductor, which would give the necessary local suppression of superconductivity needed for flux pinning.  This is probably true for the cubic phase as well, since no secondary superconducting transitions were indicated in \cite{He}.  The volume fraction of pinning sites must also be substantial, $\sim\!\! 20$\% \cite{CMb}, to provide a number density of pinning sites comparable to the flux-line density.  Thus, the present experiment indicates that there is a large fraction of precipitates present in MgCNi$_3$, which should affect its normal-state and superconducting properties.  

In particular, we observed a large slope $\mu_0 dH_{c2}/dT$ of $-1.86$~T/K at $T_c$, similar to the slope found in \cite{Li}.  This indicates that MgCNi$_3$ is a dirty-limit superconductor.  We speculate that this is a direct consequence of the electron scattering by the precipitates.  Moreover, since excess C is needed to obtain high $T_c$, graphite precipitates may be an inevitable part of the fabrication process.  It is also likely that the cubic nanoprecipitates result from the fabrication process, and may be found in MgCNi$_3$ samples used in other experiments so far.  If analogs of MgCNi$_3$ can be made with higher $T_c$ while retaining the nanostructure, these may have technologically important upper critical field values.

An interesting and perhaps very important result is the observation of strong core pinning by intragranular pinning sites in a polycrystalline intermetallic superconductor.  This combination is extremely rare; generally grain-boundary pinning is dominant, and pinning by e.g.\ intragranular structural defects only takes over when the grain sizes is very large \cite{Bonney} or when the flux lattice is very soft \cite{Kes}.  Further, these excellent flux-pinning properties are found together with high $H_{c2}$ (relative to $T_c$).  Of all the superconducting materials, perhaps only Nb-Ti alloys exhibit such a unique combination of pinning and high-field performance, and partly for this reason (ductility is the other) Nb-Ti alloys have been the mainstay of magnet technology since the 1960s.  For example, the standard expression $H_{c2}(0) = 3110 \rho\gamma T_c [{\rm tesla}]$ \cite{dirty} suggests upper critical fields of 20~T at low temperature could be obtained if $T_c$ were doubled in a doped compound with comparable resistivity $\rho$ and electronic specific heat coefficient $\gamma$ to that found in MgCNi$_3$.  Therefore, analogs of MgCNi$_3$ which have higher $T_c$ might be extremely valuable for magnet applications.

In conclusion, we have examined the flux-pinning properties and nanostructure of MgCNi$_3$.  The flux-pinning results are suggestive of a transition from pinning by grain boundaries at temperatures near $T_c$ to core pinning by 10 to 20\% of nanometer-scale precipitates at low temperatures.  Transmission electron microscopy indicated the presence of graphite and a cubic phase related to MgCNi$_3$, possibly a carbon deficient version of the same compound, with 2 to 5~nm size and substantial volume fraction.  Since the precipitates can form because of processing with excess carbon and Mg loss, scattering due to the nanostructure may be the cause of dirty-limit behavior seen so far.  If higher critical temperature can be obtained in analogs of MgCNi$_3$ which retain the nanostructure, these might exhibit technologically important upper critical field values.  Moreover, MgCNi$_3$ exhibits the unique combination of high $H_{c2}$ and strong core pinning by the nanostructure, which is not found in other intermetallic superconductors.  This makes analogs of MgCNi$_3$ potentially important for magnet technology.

This  work was supported by grants from the US National Science Foundation and the US Department of Energy.  We would like to acknowledge helpful discussions with A. Polyanskii.


\end{document}